\documentclass[12pt]{article}
\usepackage{graphicx}
\usepackage{amsmath}
\usepackage[sort&compress,numbers]{natbib}
\usepackage{doi}
\usepackage{hyperref}
\usepackage{xspace}
\usepackage[utf8]{inputenc}
\usepackage{url}
\newcommand\pubnumber{ }
\newcommand\pubdate{\today}

\def\nmsu{New Mexico State University, Physics Department\\
MSC 3D, Box 30001, Las Cruces, NM 88003, USA}
\def\support{\footnote{Supported by the Office of Nuclear Physics in the Office of Science
    of the Department of Energy.}}

\def\Title#1{\begin{center} {\Large #1 } \end{center}}
\def\Author#1{\begin{center}{ \sc #1} \end{center}}
\def\Address#1{\begin{center}{ \it #1} \end{center}}

\newcommand\pubblock{\rightline{\begin{tabular}{l} \pubnumber\\
         \pubdate  \end{tabular}}}
\newenvironment{Abstract}{\begin{quotation}  }{\end{quotation}}
\newenvironment{Presented}{\begin{quotation} \begin{center} 
             PRESENTED AT\end{center}\bigskip 
      \begin{center}\begin{large}}{\end{large}\end{center} \end{quotation}}

\newcommand{\genie}{{\sc Genie}\xspace}
\newcommand{\neut}{{\sc Neut}\xspace}
\newcommand{\geant}{{\sc Geant}4\xspace}
\newcommand{\corsika}{{\sc Corsika}\xspace}
\newcommand{\dune}{{\sc Dune}\xspace}

\def\beq{\begin{equation}}
\def\eeq#1{\label{#1}\end{equation}}
\def\eeqn{\end{equation}}

\def\beqa{\begin{eqnarray}}
\def\eeqa#1{\label{#1}\end{eqnarray}}
\def\eeqan{\end{eqnarray}}

\let\bar=\overbar

\def\Dslash{\not{\hbox{\kern-4pt $D$}}}
\def\dslash{\not{\hbox{\kern-2pt $\del$}}}

\def\msb{{\bar{\ssstyle M \kern -1pt S}}}

\begin{document}
\begin{titlepage}
\pubblock

\vfill
\Title{Neutrino Scattering Studies in MicroBooNE}
\vfill
\Author{ Vassili Papavassiliou\support\\
for the MicroBooNE Collaboration}
\Address{\nmsu}
\vfill
\begin{Abstract}
  A good understanding of the cross sections for neutrino interactions with nucleons
  and nuclei is crucial for neutrino oscillation studies, in addition to providing a
  tool for the exploration of nucleon and nuclear structures. The MicroBooNE liquid-argon
  time-projection-chamber (LArTPC) experiment has been taking neutrino data with the Booster
  Neutrino Beam at Fermilab since 2015. The LArTPC capabilities in track reconstruction,
  energy measurement, and particle identification allow us to probe interesting regions of
  neutrino-argon scattering cross sections and to probe the quark composition of the nucleon
  and test models of nuclear structure and final-state interactions. We present the current
  status of several on-going MicroBooNE cross section analyses, as well as plans for future
  measurements. 
\end{Abstract}
\vfill
\begin{Presented}
Thirteenth Conference on the Intersections of Particle and Nuclear Physics \\
Palm Springs, CA, May 29 -- June 3, 2018
\end{Presented}
\vfill
\end{titlepage}
\def\thefootnote{\fnsymbol{footnote}}
\setcounter{footnote}{0}

\section{Introduction}
The MicroBooNE experiment was designed to explore the ``low-energy excess'' (LEE)
observed~\cite{AguilarArevalo:2008rc,AguilarArevalo:2010wv} by the MiniBooNE experiment,
recently updated~\cite{Aguilar-Arevalo:2018gpe} with data from several more
years of running. MiniBooNE, a mineral-oil, Cerenkov detector, is observing a signal
of events consisting of electron-like, Cerenkov rings almost 5$\sigma$ above the
expectations from all known sources in the (predominantly muon-neutrino) Fermilab
Booster Neutrino Beam (BNB). The results may be interpreted as $\nu_\mu\rightarrow\nu_e$
oscillations involving a fourth, sterile type of neutrino; however, the energy-
dependence of the signal is not perfectly consistent with oscillations. Alternatively,
some unknown source of photons could be responsible for the signal through photon
pair conversion, as the Cerenkov detector cannot distinguish between rings from single
electrons and from nearly-collinear, $e^+e^-$ pairs, leaving the interpretation in
terms of sterile neutrinos in doubt. The technology employed by MicroBooNE, a liquid-
argon, time-projection chamber (LArTPC) can separate the two event topologies, considering
that an $e^+e^-$ pair will produce twice the amount of ionization per unit length as a
single electron track with the same total energy.

Studies of neutrino oscillations require a good understanding of neutrino scattering
on nuclear targets. Cross sections and particle-production features in scattering of
neutrinos from argon are very poorly known and models used in neutrino generators for
Monte Carlo simulations can vary greatly. Among the generators commonly used are
\genie~\cite{Andreopoulos:2009rq} (used as the default by MicroBooNE),
NuWro~\cite{Juszczak:2003zw}, \neut~\cite{Hayato:2009zz}, and GiBUU~\cite{Buss:2011mx},
and various ``tunes,''
involving different parameters, may be used with any particular generator. These
tunes and the generators themselves are constantly evolving as additional data from
neutrino experiments become available. MicroBooNE aims to measure cross sections
and perform detailed studies of particles produced in $\nu\text{-Ar}$ interactions
which will be of use, not only for this experiment, but also for future experiments
of the Short-Baseline Neutrino (SBN) program~\cite{Antonello:2015lea}, as well as the
(higher-energy) Long-Baseline Neutrino (LBN) program with the DUNE
detector~\cite{Acciarri:2015uup}. MicroBooNE has been collecting
beam data since October, 2015 and the experimental run is continuing. Both the BNB,
with peak neutrino energy at around 800~MeV and,
off-axis, the higher-energy, NuMI beam, are seen by the MicroBooNE detector. In the
following, some early, mostly preliminary, results from neutrino scattering on argon
are presented.

\section{The Experiment}
\begin{figure}[htb]
\centering
\includegraphics[width=0.8\textwidth]{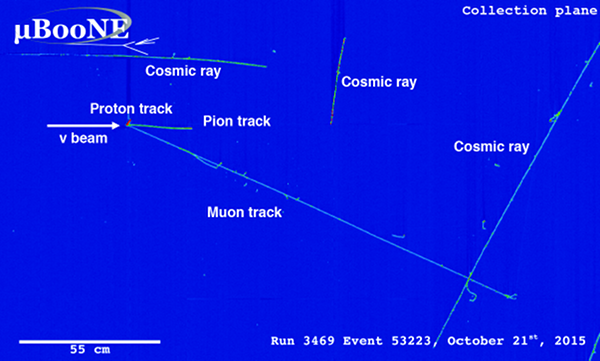}
\caption{Charged-current event showing a muon, proton, and pion tracks, with several
  cosmic-ray tracks superimposed. Tracks are labeled; the direction of the incoming
  neutrino is also indicated.}
\label{fig:CCevent}
\end{figure}

The MicroBooNE detector is described in detail in~\cite{Acciarri:2016smi}. A cryostat
is filled with 170 tons of liquid argon, of which 84 tons are contained within the
active volume of the TPC. The TPC is approximately 2.3~m tall by 2.6~m wide (the drift
direction) by 10.4~m long (along the beam direction). The anode consists of two induction
planes with wires at $\pm 60^\circ$ and a collection plane with vertical wires, for a
total of 8,256 wires. It uses a drift voltage of 70~kV for a field of 273~V/cm,
resulting in a maximum drift time of 2.3~ms. A light-collection
system~\cite{Katori:2013wqa} comprises 32,
8-inch, photomultiplier tubes with their surfaces coated with tetraphenyl butadiene,
in order to convert the VUV, 128-nm, Ar scintillation light into the visible range.
The timing provided by the fast, 6-ns component of the scintillation light plays an
important role in mitigating the effects of cosmic-ray backgrounds, as the detector
is situated just below ground level with no significant overburden for shielding.
Fig.~\ref{fig:CCevent} presents a partial event display, showing a charged-current
event with several cosmic-ray tracks overlaid; between five and ten cosmic rays
cross the detector during the 4.8~ms readout window and the scintillation light
allows tagging tracks that are coincident with the 1.6-$\mu$s beam spill. The light
is also used for triggering, as most of the beam spills do not result in any neutrino
interaction in the detector.

\begin{figure}[htb]
\centering
\includegraphics[width=0.95\textwidth]{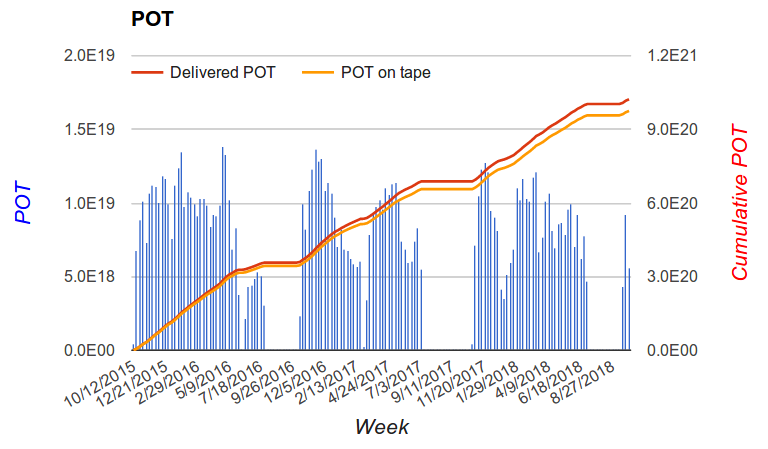}
\caption{Protons on target, delivered and recorded, from the Booster Neutrino Beam
  since the start of the MicroBooNE experimental run. Weekly fluxes (left axis) and
  the integrated flux (right axis) are shown.}
\label{fig:pot}
\end{figure}

As of this writing, the experiment has received $10^{21}$ protons on target (POT)
from the Booster beam, with 97.5\% recorded on tape.
Fig.~\ref{fig:pot} shows the weekly and integrated fluxes.

\section{Simulations}
MicroBooNE uses the \genie as the default neutrino event generator. Two tunes (sets
of parameters) have been used for comparisons with data: the default, with an empirical
meson-exchange-current (MEC) model added (the ``Dytman Model''), and an alternative
model using a theoretical MEC calculation. The main ingredients are shown in
Table~\ref{tab:Tunes}. In one of the analyses presented below, of charged-particle
multiplicities, an additional model, based on the default \genie with the addition of
a Transverse-Enhancement Model \cite{Bodek:2011ps}, was
also used for comparing with the data. Finally, the NuWro generator will also be used
in some future studies as an alternative.

\begin{table}[ht]
\begin{center}
  \begin{tabular}{|l||l|l|}
    \hline
    Tune & Default + Emp.\ MEC & Alternative\\
    \hline
    Nuclear model & Relativistic Fermi gas \cite{Bodek:1981wr}
    & Local Fermi gas \cite{Nieves:2011pp,Gran:2013kda}\\
    Quasielastic model & Llewellyn-Smith \cite{LlewellynSmith:1971uhs}
    & Nieves \cite{Nieves:2011pp,Gran:2013kda}\\
    MEC model & Empirical \cite{Katori:2013eoa}
    & Valencia \cite{Nieves:2011pp,Gran:2013kda}\\
    Resonance model & Rein-Sehgal \cite{Rein:1980wg}
    & Berger-Sehgal \cite{Berger:2008xs}\\
    Final-state interactions & hA \cite{Andreopoulos:2015wxa}
    & hA2012 \cite{Andreopoulos:2015wxa}\\
    \hline
  \end{tabular}
  \caption{Main \genie tunes used in simulations. Both are based on \genie v.2.12.2.}
  \label{tab:Tunes}
\end{center}
\end{table}

Detector simulation was based on the \geant~\cite{Allison:2016lfl,Allison:2006ve}
package. The cosmic-ray background is estimated using either the
\corsika~\cite{Heck:1998vt,Engel:2018akg} generator or off-beam data to overlay
cosmic-ray tracks on Monte-Carlo-generated, neutrino-induced events. 

\section{Highlights}
Here we present a sample of preliminary results on various processes. They are based
on a subset of data from the first experimental run (``Run 1'') beween October, 2015
and July, 2016, with the Booster beam.

\subsection{Charged-Current, Inclusive Cross Sections}
This inclusive process is sensitive to not only single-nucleon knockout
(one-particle-one-hole, or ``$1p1h$,'') processes, but also to multi-nucleon
correlations, such as MEC, mentioned above, and also possible short-range
correlations (SRC)~\cite{Fomin:2017ydn}. These two-particle-two-hole (``$2p2h$'')
processes may
have been responsible for the disagreement between the results of MiniBooNE,
which could not detect low-energy, final-state protons below the Cerenkov
threshold, and higher-energy neutrino experiments which measured cross sections
with the requirement of a single final-state proton~\cite{Formaggio:2013kya}.
Measurement of the inclusive cross section allows a more direct comparison with
predictions based on nuclear models.

Flux-integrated and single-differential cross sections for CC inclusive,
$^\mathrm{nat}\mathrm{Ar}(\nu_\mu,\mu^-)$ scattering were measured from six months of
data from Run~1, with a total of $\sim 1.6\times 10^{20}$~POT~\cite{Note:1045}.
After reconstruction and removal of cosmic-ray tracks, events were selected
requiring one muon with any number (including zero) of additional tracks. A spatial
match between an optical flash during the 1.6-$\mu$s beam window and the tracks in
the event was required. Muons were identified using a classifier, a Support Vector
Machine algorithm, in the 2D space of track length vs.\ the average charge deposited
per unit length (truncated-mean $dQ/dx$), corresponding to the average energy loss,
making use of the excellent particle-identification capabilities of the detector.
A clear separation of muons and protons was observed. Muon momentum was then
obtained using a method based on multiple Coulomb scattering (MCS), which can
be used for both contained and exiting tracks. For the former, the momentum can
also be obtained from the total range, which allows for a cross check for validating
the MCS-based method.

\begin{figure}[htb]
\centering
\includegraphics[width=0.95\textwidth]{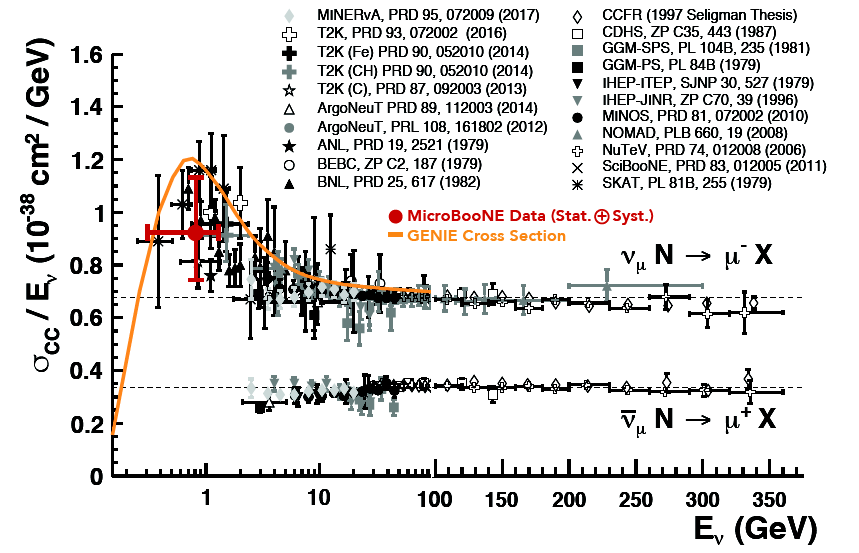}
\caption{Preliminary MicroBooNE result on the charged-current, inclusive,
  flux-integrated cross section (red dot), compared with the world data as a
  function of neutrino energy. The only other data on argon are the three
  points from the ArgoNeuT experiment (open triangles and gray dot). The \genie
  input cross section is also shown as a spline (orange curve).}
\label{fig:CCincl}
\end{figure}

Fig.~\ref{fig:CCincl} shows the inclusive cross section result from MicroBooNE
superimposed on results from neutrino and antineutrino experiments on a variety
of targets. We note that only three of these data points, from the ArgoNeuT
experiment, are on argon~\cite{Anderson:2011ce,Acciarri:2014isz}. The cross section
used as input in \genie is also displayed as a smooth curve. We find
\begin{equation}
  \sigma = 0.576 \pm 0.011\,\mathrm{(stat)}\pm0.185\,\mathrm{(syst)}
\end{equation}
The result is systematics-limited, with uncertainties in the beam flux and the
detector response being the two largest contributors.

\begin{figure}[htb]
\centering
\includegraphics[width=0.8\textwidth]{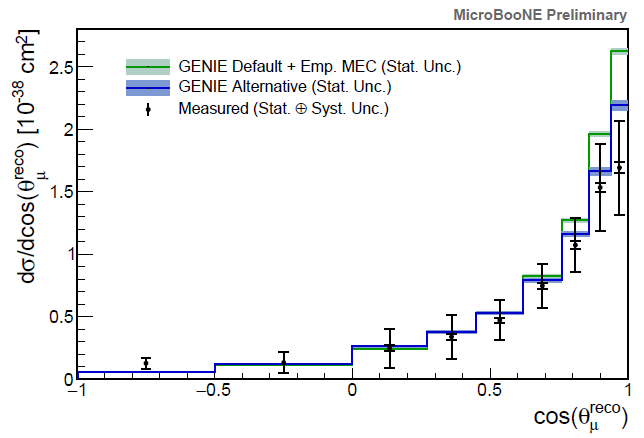}
\caption{Preliminary MicroBooNE results on the charged-current, inclusive,
  differential cross section $d\sigma/d\cos\theta_\mu$ (black dots) with statistical
  and total errors (inner and outer error bars, respectively). The two \genie
  tunes discribed in the text are also shown with statistical uncertainties,
  from Monte Carlo statistics, as the green (``Default + Emp.\ MEC'') and blue
  (``Alternative'') bands.}
\label{fig:CCinclTheta}
\end{figure}

Single-differential cross sections were also measured, as a function of the
reonstructed muon momentum and angle. Fig.~\ref{fig:CCinclTheta} shows the
cross section vs.\ $\cos\theta_\mu$ with statistical and systematic errors,
compared with the two \genie models described earlier. Both tunes overpredict
the cross section at forward angles, although the ``Alternative'' model is
closer to the data. Work continues on extracting double-differential cross
sections in bins of reconstructed $p_\mu$ and $\theta_\mu$.

\subsection{Charged-Particle Multiplicity Distributions}
Further comparisons with \genie predictions were made possible using events
with one, fully-contained, final-state muon and any number of additional,
charged tracks~\cite{Adams:2018fud}.
The charged-particle multiplicity distributions were compared
with predictions from three \genie models: the default; the default with the
empirical MEC as above; and an additional model with the \genie default
augmented by a Transverse Enhancement Model (TEM). The last two implement
$2p2h$ final-state excitations not included in the default \genie. In
addition to the multiplicity distributions, detailed data-Monte Carlo
comparisons were made of distributions in terms of a number of
kinematic variables for fixed multiplicity. Both multiplicity and
kinematic-variable distributions are sensitive to not only the nuclear
model but also the model for final-state interactions.

\begin{figure}[htb]
\centering
\includegraphics[width=0.95\textwidth]{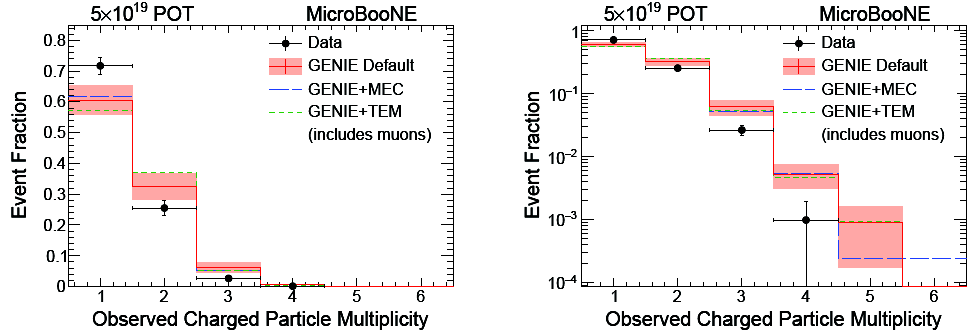}
\caption{Charged-particle multiplicities in charged-current events
  compared to the default \genie model and two models where the default
  was suplemented with either empirical MEC or the TEM (see text); in
  linear (left) or logarithmic (right) vertical scale.}
\label{fig:multiplicity}
\end{figure}

Kinematic-variable distributions at fixed multiplicity were in general
in good agreement with Monte-Carlo predictions; however, the data showed
lower average multiplicities overall compared to the simulations.
Fig.~\ref{fig:multiplicity} shows the charged-track, multiplicity
distribution from data acquired with an integrated beam flux corresponding
to $5\times 10^{19}$~POT, compared to the three \genie models. Cosmic-ray
subtraction, after event selection, was based on a \corsika simulation.
A larger fraction of events have
observed multiplicity one, while the fractions with multiplicities two
or higher are all lower than predicitions from \genie.

\subsection{Charged-Current $\pi^0$ Production}
Neutral-pion production in neutral-current interactions can be a significant
background to the oscillation searches. Showers from the two photons from the
$\pi^0$ decay can be merged and be reconstructed as a single photon, which
in some cases may be mistaken for an electron. This can cause a
$(\nu_\mu,\nu_\mu\pi^0)$ event to be misreconstructed as $(\nu_e,e)$, where
the $\nu_e$ is interpreted as signature of an oscillation signal.

\begin{figure}[htb]
\centering
\includegraphics[width=0.8\textwidth]{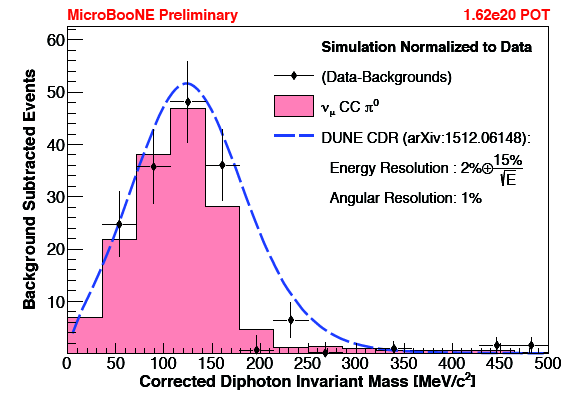}
\caption{Diphoton invariant mass for charged-current, $\pi^0$-production
  candidate events. Data after background subtraction are shown as black
  dots. The histogram shows the Monte-Carlo prediction for $\nu_\mu$ CC
  $\pi^0$ production; the expected mass resolution for \dune is also shown
  as a smooth curve.}
\label{fig:pi0mass}
\end{figure}

The performance of the shower-reconstruction algorithms can be studied and a
quantitative estimate of the above background can be obtained using the more
easily identifiable charged-current, neutral-pion production: events with a
muon and electromagnetic showers in the final state as the signature. The
shower-energy resolution can be obtained as well. Fig.~\ref{fig:pi0mass} shows
preliminary results on the reconstructed diphoton mass, after background
subtraction, from data corresponding to $1.62\times 10^{20}$ POT~\cite{Note:1032}.
A peak consistent with the $\pi^0$ mass is clearly visible. The mass resolution
is very competitive with the projected resolution for \dune, based on the design
energy and angular resolution~\cite{Acciarri:2015uup}, also shown on the plot.
The total, flux-integrated cross section for this process is measured to be
$\sigma^{\nu CC \pi^0}=[1.94\pm0.16\mathrm{ (stat)}\pm0.60\mathrm{ (syst)}]
\times10^{-38}\mathrm{cm}^2$.

\subsection{Neutral-Current, Elastic Scattering}
The neutral-current, elastic (NCE) scattering process Ar$(\nu,\nu p)$ is also
very useful for oscillation studies, as it can be employed in searches for
sterile neutrinos through flux disappearence: the process is insensitive to
(active) neutrino flavor. In addition, it is an important, comlementary, tool
for studies of the nucleon structure. In particular, it can serve as a probe
of the contribution of the strange quarks to the spin of the
nucleon~\cite{Garvey:1992cg}, usually denoted as $\Delta s$ and measured,
with various model-dependent
assumptions, in deep-inelastic, lepton-nucleon scattering (DIS). Experiments
over the last thirty years point to a (surprisingly) negative
$\Delta s$~\cite{Aidala:2012mv}.
However, this has not been confirmed by studies of semi-inclusive DIS
production and the overall model-dependence demands an independent probe
of this quantity, which affects predictions in searches for certain types
of dark matter~\cite{Ellis:2018dmb}, as well as being one of the major
unknown quantities of the nucleon structure.

The NCE cross section is sensitive to $\Delta s$ and can help extract it when
combined with data from CC neutrino scattering and charged-lepton elastic
scattering. Measurements at low $Q^2$, (four-momentum transfer squared), are
needed. The signature for the NCE events is a single, proton
track and at low $Q^2$ the track will be short.

A method has been devised to identify proton tracks
based on a machine-learning algorithm, gradient-boosting decision trees,
using a number of reconstructed event variables. Then a logistic-regression
model is used to combine the information from the decision trees with other
event features to obtain an enriched sample of NCE events. The final step
involves a comparison with the \genie prediction for this process and
a fit that will optimize the form factor that \genie uses as input;
a Markov Chain Monte Carlo has been written for this purpose. This
work is in progress~\cite{Woodruff:2018}.

\bibliography{scattering}




\end{document}